\documentclass[aps,showpacs,manuscript,12pt]{revtex4}
\usepackage{amssymb}
\usepackage{amsmath}
\usepackage{graphicx}

\begin{document}
\title{\textbf{Inverse kinetic theory for incompressible
thermofluids$^{\S }$ }}
\author{C. Cremaschini$^{a}$ and and M.
 Tessarotto$^{b,c}$}
\affiliation{\ $^{a}$Department of Astronomy, University of
Trieste, Italy, $^{b}$Department of Mathematics and Informatics,
University of Trieste, Italy, $^{c}$Consortium of
Magneto-fluid-dynamics, University of Trieste, Italy}
\begin{abstract}
An interesting issue in fluid dynamics is represented by the
possible existence of inverse kinetic theories (IKT) which are
able to deliver, in a suitable sense, the complete set of fluid
equations which are associated to a prescribed fluid. \ From the
mathematical viewpoint this involves the formal description of a
fluid by means of a classical dynamical system which advances in
time the relevant fluid fields. The possibility of defining an IKT
for the 3D incompressible Navier-Stokes equations (INSE), recently
investigated (Ellero \textit{et al}, 2004-2007) raises the
interesting question whether the theory can be applied also to
thermofluids, in such a way to satisfy also the second principle
of thermodynamics. The goal of this paper is to prove that such a
generalization is actually possible, by means of a suitable
\textit{extended phase-space formulation}. We consider, as a
reference test, the case of non-isentropic incompressible
thermofluids, whose dynamics is described by the Fourier and the
incompressible Navier-Stokes equations, the latter subject to the
conditions of validity of the Boussinesq approximation.
\end{abstract}
\date{\today }
\maketitle



\section{Introduction}

A remarkable aspect of fluid dynamics is related to the
construction of inverse kinetic theories (IKT) for hydrodynamic
equations in which the fluid fields are identified with suitable
moments of an appropriate kinetic probability distribution. \
Recently the topic has been the subject of theoretical
investigations on the incompressible Navier-Stokes (N-S)
equations (INSE) \cite%
{Ellero2004,Ellero2005,Tessarotto2006,Tessarotto2006b,Tessarotto2007}.
\ The importance of the IKT-approach goes beyond the academic
interest. In fact, fluid equations represent usually a mixture of
hyperbolic and elliptic pde's, which are extremely hard to study
both analytically and numerically. As such, their investigation
represents a challenge both for mathematical analysis and for
computational fluid dynamics. For this reason in the past
alternative approaches, based on asymptotic kinetic theories, have
been devised which permit to advance in time the fluid fields, to
be determined in terms of suitable moments of an appropriate
kinetic distribution function. An example is provided by kinetic
theories for incompressible fluids which adopt the so-called
Lattice-Boltzmann approach \cite{Succi} (see also related
discussion in Refs. \cite{Tessarotto2008a,Fonda2008a}). These
methods, which approximate the fluid equations only in an
asymptotic sense, are based on the introduction of suitably
modified (fluid) equations which permit to advance in time the
fluid fields only in an approximate sense. In particular,
typically, their modified fluid equations actually describe
weakly-compressible fluids. \ The discovery of IKT
\cite{Ellero2000} provides, however, a new starting point for the
theoretical and numerical investigation of hydrodynamic equations,
since it does not require any modification of the exact fluid
equations, in particular it holds for strong solutions, and
permits to advance in time exactly the fluid fields by means
of a suitable kinetic distribution function $f(\mathbf{x,}t)$. Here $\mathbf{%
x}$ is the state vector
$\mathbf{x}=(\mathbf{r}_{1}\mathbf{,v}_{1}),$ where respectively
$\mathbf{r}_{1}$ and $\mathbf{v}_{1}$ denote the corresponding
"configuration" and "velocity" vectors, and $\Gamma $ is the
phase-space spanned by $\mathbf{x.}$ In the sequel we shall assume
that $\Gamma $ is an \emph{extended phase-space, }i.e., it has a
dimension $2n$ with $n>3$. This
is achieved introducing a phase-space classical dynamical system%
\begin{equation}
\mathbf{x}_{o}\rightarrow \mathbf{x}(t)=T_{t,t_{o}}\mathbf{x}_{o},
\label{classical dynamical system}
\end{equation}%
which uniquely advances in time the fluid fields by means of an
appropriate evolution operator $T_{t,t_{o}}$
\cite{Tessarotto2006,Tessarotto2006b}. This
is assumed to be generated by a suitably smooth vector field $\mathbf{X}(%
\mathbf{x},t),$
\begin{eqnarray}
\frac{d}{dt}\mathbf{x} &=&\mathbf{X}(\mathbf{x},t), \\
\mathbf{x}(t_{o}) &=&\mathbf{x}_{o},
\end{eqnarray}%
Therefore, introducing the corresponding microscopic distribution function $%
f(\mathbf{x,}t),$ it fulfills necessarily in $\Gamma $ the
differential Liouville equation
\begin{equation}
Lf(\mathbf{x},t)=0,  \label{Liouville}
\end{equation}%
where $L$ denotes the Liouville streaming operator $Lf\equiv
\frac{\partial
}{\partial t}f+\frac{\partial }{\partial \mathbf{x}}\cdot \left\{ \mathbf{X}(%
\mathbf{x},t)f\right\} $. This equation, which may be interpreted
as a Vlasov-type inverse kinetic equation (IKE), can in principle
be defined in such a way to satisfy appropriate constraint
equations. In particular, thanks to the arbitrariness of the
dynamical systems (\ref{classical
dynamical system}) [i.e., \ the arbitrariness of $\mathbf{X}(\mathbf{x},t)$%
], the velocity moments of $f(\mathbf{x,}t)$ might be identified -
in principle - so that suitable velocity moments\ (of $f$)
coincide with the relevant fluid fields characterizing a
prescribed classical fluid. For
example, as in Refs. \cite%
{Ellero2004,Ellero2005,Tessarotto2006,Tessarotto2006b,Tessarotto2007}
one can impose that the first velocity-moment coincides with the
fluid mass
density, i.e., there results $\rho =\int\limits_{%
\mathbb{R}
^{n}}d\mathbf{v}_{1}f(\mathbf{x,}t).$

An interesting issue is whether the theory can be applied also to
thermofluids. Such a generalization, as shown in an accompanying paper \cite%
{Tessarotto2008a}, is actually non-unique. In particular, the goal
of this
paper is to prove that an IKT can be achieved by means of a suitable \emph{%
extended phase-space formulation}. We consider, as a reference
problem, the case of incompressible thermofluids subject, for
greater generality, to the condition of non-isentropic flow. For
definiteness, {we shall assume that the relevant fluids }$\left\{
\rho =\rho _{o}>0,\mathbf{V},p\geq 0,T>0,S_{T}\right\} ,${\ i.e.,
respectively the (constant) mass density, fluid velocity,
pressure, temperature and entropy describing the fluid, are
defined in an appropriate existence domain. In particular if
$\Omega $ is an open connected subset of }$R^{3}$ {(denoted as
configuration domain; with prescribed fixed boundary }$\delta
\Omega $ and{\ closure }$\overline{\Omega }$ defined as the set
where the mass density is a constant $\rho _{o}>0${) and }$I${\ a
finite time interval $I=$}$\left] {t_{0},t_{1}}\right[ ${(with
closure }$\overline{{I}}${$=[t_{0},t_{1}]$), }we assume that the
fluid fields $\left\{ p,\mathbf{V},T\right\} $ are continuous in
$\overline{\Omega }\times \overline{{I}},$ satisfy suitable
initial and boundary conditions
respectively at $t=t_{o}$ and on $\delta \Omega ,$ while in the open set $%
\Omega \times ${$I$ they satisfy the so-called non-isentropic and
incompressible Navier-Stokes-Fourier equations (INSFE), i.e.,} {\
\begin{eqnarray}
&&\left. \nabla \cdot \mathbf{V}=0,\right.   \label{1} \\
&&\left. \frac{\partial }{\partial t}\mathbf{V}+\mathbf{V}\cdot
\nabla \mathbf{V}+\frac{1}{\rho _{o}}\left[ \nabla
p-\mathbf{f}\right] -\nu \nabla
^{2}\mathbf{V}=0,\right.   \label{2} \\
&&\left. \frac{\partial T}{\partial t}+\mathbf{V\cdot }\nabla
T=\chi \nabla
^{2}T+\frac{\nu }{2c_{p}}\left( \frac{\partial V_{i}}{\partial x_{k}}+\frac{%
\partial V_{k}}{\partial x_{i}}\right) ^{2}+\frac{1}{\rho _{o}c_{p}}%
J,\right.   \label{3} \\
&&\left. \frac{\partial }{\partial t}S_{T}\geq 0.\right.
\label{4}
\end{eqnarray}%
}Here{\ the notation is standard. Thus, Eq.(\ref{1}) denotes the
so-called \emph{isochoricity condition}, while Eq.(\ref{2}) is the
\emph{Navier-Stokes equation} in the Boussinesq approximation.
Hence, in such a case the force density }$\mathbf{f}$ {reads
}$\mathbf{f}=\rho _{o}\mathbf{g}\left( 1-k_{\rho }T\right)
+\mathbf{f}_{1},$ where the first term represents the
(temperature-dependent) gravitational force density, while the second\ one ($%
\mathbf{f}_{1}$) the action of a possible non-gravitational
externally-produced force. Moreover:

\begin{itemize}
\item Eq.(\ref{3}) is the \emph{Fourier equation} for the temperature $T,$
to be assumed strictly positive in $\overline{\Omega }\times \overline{{I}}$%
, with $J$ the quantity of heat generated by external sources per
unit volume and unit time (for example, Joule heating). Thus for
an isolated fluid there results by definition $J\equiv 0$ in
$\overline{\Omega }\times \overline{{I}}.$

\item Eq.(\ref{4}) defines the so-called \emph{2nd} \emph{principle} for the
thermodynamic entropy $S_{T}.$ For its validity in the sequel we
shall assume that there results either everywhere in
$\overline{\Omega }\times
\overline{{I}},$ $J\equiv 0$ (\emph{thermally-isolated thermofluid}) or%
\begin{equation}
\int_{\Omega }d\mathbf{r}\left( \chi \nabla ^{2}T+\frac{1}{\rho _{o}c_{p}}%
J\right) \geq 0  \label{fluid with external heating}
\end{equation}%
(\emph{externally heated thermofluid}).

\item In these equations $\mathbf{g,}$ $k_{\rho },\nu ${$,$ }$\chi $ and $%
c_{p}$ are all real constants which denote respectively the local
acceleration of gravity, the density thermal-dilatation
coefficient, the kinematic viscosity, the thermometric
conductivity and the specific heat at
constant pressure. Thus, by taking the divergence of the N-S equation (\ref%
{1}), there it follows the Poisson equation for the fluid pressure
$p$ which reads
\begin{equation}
\nabla ^{2}p=-\rho _{o}\nabla \cdot \left( \mathbf{V}\cdot \nabla \mathbf{V}%
\right) +\nabla \cdot \mathbf{f},  \label{5}
\end{equation}%
with $p$ to be assumed non negative and bounded in
$\overline{\Omega }\times \overline{{I;}}$
\end{itemize}

Finally, it is assumed that:

\begin{itemize}
\item  Eqs. ({\ref{1}})-(\ref{3}) {satify a suitable initial-boundary value
problem\ (INSFE problem) and that a smooth (strong) solution
exists for the fluid fields }$\left\{ \rho =\rho
_{o}>o,\mathbf{V},p\geq 0,T>0\right\} ;$

\item  the entropy functional $S_{T}$ can be defined so that it satisfies
the 2nd principle. \
\end{itemize}

\section{Extended phase-space IKT for incompressible thermofluids}

Here we intend to show that an IKT for INSFE can be reached by
introducing of a suitable \emph{extended phase-space formulation},
based on a generalization of the IKT developed previously for the
incompressible
Navier-Stokes equations \cite%
{Ellero2004,Ellero2005,Tessarotto2006,Tessarotto2006b,Tessarotto2007}.
For definiteness, let us introduce the notations

\begin{eqnarray}
\mathbf{r}_{1} &\mathbf{=}&\left( \mathbf{r,}\vartheta \right) ,  \notag \\
\mathbf{v}_{1} &\mathbf{=}&\left( \mathbf{v,}w\right) , \\
\mathbf{X} &=&\left\{
\mathbf{v}_{1},\mathbf{F}_{1}(\mathbf{x},t)\right\} ,
\notag \\
\mathbf{F}_{1}(\mathbf{x},t) &=&\left\{ \mathbf{F}(\mathbf{x},t),H(\mathbf{x}%
,t)\right\}  \notag
\end{eqnarray}%
where the vectors $\mathbf{r}$ and $\mathbf{v}$ span,
respectively, the whole configuration domain of the fluid
($\overline{\Omega }$) and the
3-dimensional velocity space\textbf{\ }($\mathbf{%
\mathbb{R}
}^{3}$). Moreover, $\vartheta ,w\in
\mathbb{R}
$ are two additional real (hidden) variables, with $\vartheta $
denoting in particular an ignorable configuration-space variable
[both for the fluid fields and the kinetic distribution function
$f(\mathbf{x,}t)$] defined in a bounded interval $I_{\vartheta
}=\left[ \vartheta _{0},\vartheta _{1}\right] \subset
\mathbb{R}
$. The streaming operator $L$ in this case reads $L\equiv \frac{\partial }{%
\partial t}+\mathbf{v\cdot }\frac{\partial }{\partial \mathbf{r}}+w\frac{%
\partial }{\partial \vartheta }+\frac{\partial }{\partial \mathbf{v}}\cdot
\left\{ \mathbf{F}(\mathbf{x},t)\right\} +\frac{\partial }{\partial w}%
\left\{ H(\mathbf{x},t)\right\} $, where $\mathbf{F}_{1}(\mathbf{x}%
,t)=\left\{ \mathbf{F}(\mathbf{x},t),H(\mathbf{x},t)\right\} $ can
be
interpreted as a mean field force acting on a particle with state $\mathbf{x}%
=(\mathbf{r,}\vartheta ,\mathbf{v,}w).$ In the sequel we intend to
prove
that, at least in a suitable finite time-interval $I$, the fluid fields $%
\mathbf{V},p,T$ can be identified with the velocity moments $\int\limits_{%
\mathbb{R}
^{n}}d\mathbf{v}_{1}G(\mathbf{x,}t)f(\mathbf{x,}t),$ where respectively $G(%
\mathbf{x,}t)=\mathbf{v/}\rho _{o},\left( \mathbf{v-V}\right)
^{2}/3,mw^{2}/3\rho _{o}$ and $f(\mathbf{x,}t)$ is a properly
defined kinetic distribution function$.$ In addition, if the same
distribution function $f(\mathbf{x,}t)$ is strictly positive in
the whole set $\Gamma \times I$ and the statistical entropy
functional $S(f)=-\int\limits_{\Gamma
}d\mathbf{x}f(\mathbf{x,}t)\ln f(\mathbf{x,}t)$ exists for all
$t\in I,$ we
intend to show that the thermodynamic entropy can always be identified with $%
S(f),$i.e., that
\begin{equation}
S_{T}\equiv S(f).  \label{moment-2}
\end{equation}%
To reach the proof, let us first show that, by suitable definition
of the "force" fields $\mathbf{F}(\mathbf{x},t)$ and
$Q(\mathbf{x},t),$ a particular solution of the the IKE
(\ref{Liouville}) is delivered by the (extended-space) Maxwellian
distribution:
\begin{equation}
f_{M}(\mathbf{x},t)=\frac{\rho }{\pi ^{2}v_{th,p}^{3}v_{th,T}}\exp \left\{ -%
\frac{u^{2}}{v_{th,p}^{2}}-\frac{w^{2}}{v_{th,T}^{2}}\right\} .
\label{Maxwellian}
\end{equation}%
Here $\mathbf{u}=\mathbf{v}-\mathbf{V}(\mathbf{r},t)$ is the
relative
velocity, while $v_{th,p}=\sqrt{2p_{1}(\mathbf{r},t)/\rho }$ and $v_{th,T}=%
\sqrt{2T(\mathbf{r},t)/m}$ denote respectively the pressure and
temperature thermal velocities. Furthermore
$p_{1}(\mathbf{r},t)=p_{0}+p(\mathbf{r},t)$ is the kinetic
pressure. In these definitions, $p_{0}(t)$ (to be denotes as
\emph{pseudo-pressure}) is an arbitrary strictly positive and
suitably smooth function defined in $I,$ while the mass $m>0$ is
an arbitrary real constant. The following theorem can immediately
be proven:

\textbf{Theorem 1 - Local Maxwellian solution for INSFE-IKT }
\emph{Let us identify respectively the vector and scalar fields }$F(x,t)$\emph{\ and }$%
H(x,t)$\emph{\ with }%
\begin{equation}
\mathbf{F}(\mathbf{x},t;f_{M})=\mathbf{F}_{0}+\mathbf{F}_{1}
\end{equation}

\begin{equation}
H(\mathbf{x},t;f_{M})=\frac{w}{2T}K+\frac{w}{2}\mathbf{u\cdot }\frac{%
\partial }{\partial \mathbf{r}}\ln T,
\end{equation}

\emph{where }$\mathbf{F}_{0}\mathbf{,F}_{1}$\emph{\ and
}$K$\emph{\ read
respectively }%
\begin{equation}
\mathbf{F}_{0}\mathbf{(x,}t;f_{M})=\frac{1}{\rho _{o}}\mathbf{f}+\frac{1}{2}%
\mathbf{u}\cdot \nabla \mathbf{V+}\frac{1}{2}\mathbb{\nabla
}\mathbf{V\cdot u+}\nu \nabla ^{2}\mathbf{V,}
\end{equation}%
\begin{equation}
\mathbf{F}_{1}\mathbf{(x,}t;f_{M})=\frac{\mathbf{u}}{2p_{1}}A+\frac{%
v_{th}^{2}}{2}\nabla \ln p_{1}\left\{ \frac{u^{2}}{v_{th}^{2}}-\frac{3}{2}%
\right\} ,
\end{equation}%
\begin{equation}
K=\chi \nabla ^{2}T+\frac{\nu }{2c_{p}}\left( \frac{\partial
V_{i}}{\partial
x_{k}}+\frac{\partial V_{k}}{\partial x_{i}}\right) ^{2}+\frac{J}{\rho c_{p}}%
,  \label{kappa}
\end{equation}%
\emph{while, denoting }$\frac{D}{Dt}=\frac{\partial }{\partial t}+\mathbf{V}%
\boldsymbol{\cdot \nabla ,}$\emph{\ }$A$ \emph{reads:}%
\begin{equation}
A\equiv \frac{D}{Dt}p_{1}=\frac{\partial }{\partial t}p_{1}-\mathbf{V}\cdot %
\left[ \frac{\partial }{\partial t}\mathbf{V}+\mathbf{V}\cdot \nabla \mathbf{%
V}-\frac{1}{\rho _{o}}\mathbf{f}-\nu \nabla ^{2}\mathbf{V}\right]
.
\end{equation}

\emph{It follows that:}

\emph{1) the local Maxwellian distribution (\ref{Maxwellian}) is a
solution of the IKE (\ref{Liouville}) if an only if the fluid
fields }$\left\{ \rho
=\rho _{o}>0,\mathbf{V},p,T\right\} $\emph{\ satisfy the fluid equations (%
\ref{1})-(\ref{3});}

\emph{2) the velocity-moment equations obtained by taking the
weighted
velocity integrals of Eq.(\ref{Liouville}) with the weights }$%
G(x,t)=1,v/\rho _{o},\left( \mathbf{v-V}\right)
^{2}/3,w^{2}/3$\emph{\ deliver identically the same fluid
equations (\ref{1})-(\ref{3}).}\newline
\noindent PROOF - First we notice that if the fluid equations (\ref{1})-(\ref%
{3}) are satisfied identically in $\Omega \times I,$ the proof that (\ref%
{Maxwellian}) is a particular solution of the IKE
[Eq.(\ref{Liouville})] follows by direct differentiation. The
converse implication, i.e., the proof that if (\ref{Maxwellian})
is a solution of Eq.(\ref{Liouville}) then the fluid equations
(\ref{1})-(\ref{3}) are satisfied identically in $\Omega
\times I,$ follows by evaluating the velocity moments of Eq.(\ref{H-theorem}%
) for the weights $G=1,\mathbf{v,}\left( \mathbf{v-V}\right)
^{2}/3,w^{2}/3.$ In analogy to the case of isothermal fluids the
present theorem can be generalized to suitably smooth
non-Maxwellian initial distribution function \cite{Ellero2005}. In
such a case the fields $\mathbf{F}_{0}\mathbf{,F}_{1}$ and $H$
read respectively
\begin{equation}
\mathbf{F}_{0}\mathbf{(x,}t;f)=\frac{1}{\rho _{o}}\left[
\mathbf{\nabla \cdot }\underline{\underline{{\Pi
}}}-\mathbf{\nabla }p_{1}+\mathbf{f}\right] +\mathbf{u}\cdot
\nabla \mathbf{V+}\nu \nabla ^{2}\mathbf{V,} \label{F0
non-maxwellian case}
\end{equation}%
\begin{equation}
\mathbf{F}_{1}\mathbf{(x,}t;f)=\frac{1}{2}\mathbf{u}\left\{ \frac{1}{p_{1}}A%
\mathbf{+}\frac{1}{p_{1}}\mathbf{\nabla \cdot
Q}-\frac{1}{p_{1}^{2}}\left[
\mathbf{\nabla \cdot }\underline{\underline{{\Pi }}}\right] \mathbf{\cdot Q}%
\right\} +\frac{v_{th}^{2}}{2p_{1}}\mathbf{\nabla \cdot }\underline{%
\underline{{\Pi }}}\left\{
\frac{u^{2}}{v_{th}^{2}}-\frac{3}{2}\right\} , \label{F1
non-Maxwellian case}
\end{equation}%
while the scalar field $H(\mathbf{x},t;f)$ now reads%
\begin{equation}
H(\mathbf{x},t;f)=\frac{w}{2T}K_{1}+\frac{w}{2}\mathbf{u\cdot }\frac{%
\partial }{\partial \mathbf{r}}\ln T.  \label{non-maxwellian  H}
\end{equation}%
Here $K_{1}$ denotes the scalar function $K_{1}=K-\left[ {\Phi
}_{T}\cdot \nabla \ln T+\nabla \cdot \left( \mathbf{V}_{T}T\right)
-\mathbf{V\cdot \nabla }T\right] ,$ where $K$ is defined by
Eq.(\ref{kappa}) and furthermore
the moments $\mathbf{Q,}$\textbf{\ }$\underline{\underline{{\Pi }}},{\Phi }%
_{T}$ and \ $\mathbf{V}_{T}$ are defined respectively as
$\mathbf{Q}=\int d^{3}vdw\mathbf{u}\frac{u^{2}}{3}f,$
$\underline{\underline{{\Pi }}}=\int d^{3}vdw\mathbf{uu}f,$ ${\Phi
}_{T}=\int d^{3}vdw\frac{w^{2}}{2}\mathbf{u}f$
and $\mathbf{V}_{T}=\frac{1}{\rho _{o}T}\int d^{3}vdw\frac{w^{2}}{2}\mathbf{v%
}f.$

\section{H -Theorem and the 2nd principle of thermodynamics}

In this section we want to prove that, provided the pseudo-pressure $%
p_{0}(t) $ is suitably defined (i.e., is a uniquely-prescribed
function of time), an H Theorem can be established for the
statistical entropy $S\left( f\right) =-\int_{\Gamma
}d\mathbf{x}f\ln f,$ which warrants the strict positivity of the
kinetic distribution function in the whole set $\Gamma \times I.$
As a further consequence, the position (\ref{moment-2}) holds too.
For this purpose we distinguish between isothermal and
non-isothermal fluids, i.e., fluids in which the fluid temperature
is respectively a constant in the whole set \emph{\
}$\overline{\Omega }\times \overline{I},$ or not. In particular,
we intend to prove that, in the first case (isothermal fluid), a
constant H-theorem holds\ for the statistical entropy under
suitable conditions, i.e., by suitably prescribing the
pseudo-pressure. Instead, to reach an H-theorem for a
non-isothermal fluid, it is necessary to include also a suitable
prescription on the r.h.s. of the
Fiourier equation [and in particular on the scalar field $J;$ see Eq.(\ref{3}%
)], which defines the quantity of heat generated by external
sources. In both cases, the result can be proven to hold at least
in a finite time interval $I$ and for an arbitrary strictly
positive (and suitably summable) distribution function $f$. \

\textbf{Theorem 2 - H-theorem } \emph{Let us assume that:}
\emph{1) }$\Omega
$\emph{\ is a bounded subset of }$%
\mathbb{R}
^{3}$\emph{;} \emph{2) the kinetic distribution function }$f$ \emph{%
coincides identically in }$\overline{\Omega }\times \overline{I}$
\emph{\
with the local Maxwellian distribution f}$_{M}$\emph{\ defined by Eq.(\ref%
{Maxwellian}).} \emph{Furthermore, let us distinguish respectively
the cases
in which the fluid is isothermal in }$\overline{\Omega }\times \overline{I}$%
\emph{\ or not. In the first case we demand that the following
assumption is
fulfilled: } $3A)$\emph{\ the pseudo-pressure }$p_{0}(t)$\emph{\ is }$%
p_{0}(t)>0$\emph{\ determined in such a way to satisfy identically
}$\forall
t\in I$\emph{\ the constraint}%
\begin{equation}
\int_{\Omega }d\mathbf{r}\frac{1}{p_{1}}\left[ \frac{\partial }{\partial t}%
p_{1}+\mathbf{\nabla \cdot Q-}\frac{1}{p_{1}}\mathbf{\nabla }p\mathbf{\cdot Q%
}\right] =0.  \label{constraint on po-0}
\end{equation}%
\emph{Instead, for non-isothermal fluids, we require that the
following two
assumptions are satisfied (3B and 4):} \emph{3B) the pseudo-pressure }$%
p_{0}(t)$\emph{\ is }$p_{0}(t)>0$\emph{\ and satisfies identically
}$\forall
t\in I$\emph{\ the constraint}%
\begin{eqnarray}
&&\left. \frac{3}{2}\rho _{o}\int_{\Omega
}d\mathbf{r}\frac{1}{p_{1}}\left[
\frac{\partial }{\partial t}p_{1}+\mathbf{\nabla \cdot Q-}\frac{1}{p_{1}}%
\mathbf{\nabla }p\mathbf{\cdot Q}\right] -\right.
\label{constarint on po-1}
\\
&&\left. -\frac{1}{2}\rho _{o}\int_{\Omega }d\mathbf{r}\frac{1}{T}\left[ {%
\Phi }_{T}\cdot \nabla \ln T+\nabla \cdot \left( \mathbf{V}_{T}T\right) %
\right] =0\right. ;  \notag
\end{eqnarray}%
\emph{4) the quantity of heat generated by external sources
}$J,$\emph{\
either vanishes identically in }$\overline{\Omega }\times \overline{I}$\emph{%
\ (isolated fluid) or }$\forall t\in I$\emph{\ is externally
heated in the sense of the inequality (\ref{fluid with external
heating}) .}

\emph{Then it follows respectively:} \emph{A) for isothermal
fluids: the statistical entropy }$S\left( f\right) $\emph{\ is
constant, i.e., there
holds identically the constant H-theorem:}%
\begin{equation}
\frac{\partial }{\partial t}S\left( f\right) =0;
\label{constant-H-theorem}
\end{equation}
\emph{B) for non-isothermal fluids: the statistical entropy
}$S\left( f\right) $\emph{\ is a monotonic function of time, i.e.,
it holds, instead, the H-theorem}$\forall t\in I$\emph{\ }
\begin{equation}
\frac{\partial }{\partial t}S\left( f\right) \geq 0.
\label{H-theorem}
\end{equation}%
\noindent PROOF - For definiteness let us first consider an
isothermal fluid, i.e., requiring $T=const.$ in $\overline{\Omega
}\times \overline{I}.$ In this case the proof of
Eq.(\ref{constant-H-theorem}) is immediate. In fact for an
arbitrary, suitably smooth and non vanishing, non-Maxwellian
distribution the entropy production rate $\frac{\partial }{\partial t}%
S\left( f\right) $ results
\begin{eqnarray}
&&\left. \frac{\partial }{\partial t}S\left( f\right) =-\frac{\partial }{%
\partial t}\int d\mathbf{x}f\ln f=\right. \\
&=&\frac{3}{2}\rho _{o}\int_{\Omega }d\mathbf{r}\frac{1}{p_{1}}\left[ \frac{%
\partial }{\partial t}p_{1}+\mathbf{\nabla \cdot Q-}\frac{1}{p}\mathbf{%
\nabla }p\mathbf{\cdot Q}\right] ,  \notag
\end{eqnarray}%
which thanks to the constraint equation Eq.(\ref{constraint on po-0}) for $%
p_{0}(t)$ vanishes identically. In particular, in the case in
which the kinetic distribution function $f$ coincides with the
local Maxwellian distribution (\ref{Liouville}) the same equation
delivers
\begin{equation}
\frac{\partial }{\partial t}S\left( f\right) =\frac{3}{2}\rho
_{o}\int_{\Omega }d\mathbf{r}\frac{\partial }{\partial t}\ln
\left[ p_{0}(t)+p(\mathbf{r},t)\right] =0.
\end{equation}%
Let us now consider the case of a non-isothermal fluid obeying the
fluid equations (\ref{1})-(\ref{3}). \ In this case, imposing on
$p_{0}(t)$ the
constraint (\ref{constarint on po-1}) the entropy production rate reads%
\begin{equation}
\frac{\partial }{\partial t}S\left( f\right) =\frac{1}{2}\rho
_{o}\int_{\Omega }d\mathbf{r}K.
\end{equation}%
Hence, for a fluid which is either isolated or subject to external
heating, in the sense of the inequality (\ref{fluid with external
heating}), the H-theorem (\ref{H-theorem}) manifestly holds.

This result is consistent with the second principle if
thermodynamics [i.e., the inequality (\ref{4})]. As a consequence,
this enables us to specify also the thermodynamic entropy in terms
of a suitable phase-space moment of the kinetic distribution
function $f$.

\section{Concluding remarks}

A basic implication of the IKT here developed \ for INSFE is that
it has been constructed in such a way to satisfy the following
requirements:

\begin{enumerate}
\item \emph{completeness:} all fluid fields are expressed as moments of the
kinetic distribution function and all hydrodynamic equations can
be identified with suitable moment equations of IKE;

\item \emph{closure condition of moment equations:} there must exist a
subset of moments of IKE which form a complete system of
equations, to be identified with the prescribed set of
hydrodynamic equations;

\item \emph{smoothness for the fluid fields}: the fluid fields are assumed
suitably smooth so that the solution of the kinetic distribution
function exists everywhere in a suitable phase-space;

\item \emph{arbitrary initial and boundary conditions for the fluid fields}:
the initial conditions for the fluid equations are set arbitrarily
while Dirichlet boundary conditions are considered on the
boundary;

\item \emph{self-consistency}: the kinetic theory holds for arbitrary (and
suitably smooth) initial conditions for the kinetic distribution
function.

\item \emph{non-asymptotic IKE:} i.e., the correct hydrodynamic equations
must be recovered by the inverse kinetic theory independently of
any dimensionless parameter.
\end{enumerate}

The present approach has the following main features:

\begin{enumerate}
\item a suitable classical dynamical system has been constructed which
uniquely determines the evolution of the fluid fields;

\item the IKT is based on an extended phase-space formulation which relies
of the microscopic statistical description of the dynamical
system;

\item the theory satisfies the second principle of thermodynamics.
\end{enumerate}

An interesting result of the theory, relevant for the mathematical
investigation of the fluid equations, concerns the discovery of
the underlying dynamical system, i.e., the phase-space classical
dynamical system (\ref{5}). We have found that this can be
identified with a - generally non-conservative - dynamical system
advancing in time the microscopic distribution function and
generated by the kinetic equation itself. The evolution of the
fluid fields is thus determined uniquely by this dynamical system,
a result that in principle may be achieved without solving
explicitly the fluid equations themselves.

\section*{Acknowledgments}
Work developed in cooperation with the CMFD Team, Consortium for
Magneto-fluid-dynamics (Trieste University, Trieste, Italy). \
Research developed in the framework of the MIUR (Italian Ministry
of University and Research) PRIN Programme: {\it Modelli della
teoria cinetica matematica nello studio dei sistemi complessi
nelle scienze applicate}. The support of COST Action P17 (EPM,
{\it Electromagnetic Processing of Materials}) and GNFM (National
Group of Mathematical Physics) of INDAM (Italian National
Institute for Advanced Mathematics) is acknowledged.

\section*{Notice}
$^{\S }$ contributed paper at RGD26 (Kyoto, Japan, July 2008).
\newpage



\end{document}